\documentclass[11pt]{article}
\usepackage{jcappub}
\usepackage{amssymb,graphicx}
\usepackage{epstopdf}
\usepackage{amsmath,amsfonts}
\usepackage{epsfig}

\def\e{{\rm e}}

\def\d{\partial}
\def\l{\left(}
\def\r{\right)}

\newcommand{\be}{\begin{equation}}
\newcommand{\ee}{\end{equation}}
\newcommand{\bea}{\begin{eqnarray}}
\newcommand{\eea}{\end{eqnarray}}
\newcommand{\bg}{\begin{gather}}
\newcommand{\eg}{\end{gather}}
\newcommand{\bseq}{\begin{subequations}}
\newcommand{\eseq}{\end{subequations}}

\renewcommand{\tanh}{\mathop{\rm th}\nolimits}

\renewcommand{\ln}{\mathop{\rm ln}\nolimits}

\title{\boldmath
Generalized Galileons: instabilities
 \\[0.3cm] of bouncing and
Genesis cosmologies   \\[0.3cm] and modified Genesis
}
\author[a,b]{M.~Libanov}
\author[a]{S.~Mironov}
\author[a,c]{V.~Rubakov}
\affiliation[a]{
Institute for Nuclear Research of
         the Russian Academy of Sciences,\\  60th October Anniversary
  Prospect 7a, 117312 Moscow, Russia
}
\affiliation[b]{
Moscow Institute of Physics and Technology,\\
Institutskii per. 9, 141700 Dolgoprudny, Moscow Region, Russia
}
\affiliation[c]{
Department of Particle Physics and Cosmology,\\
Physics Faculty, M.V. Lomonosov Moscow State University,\\ Vorobjevy Gory,
119991, Moscow, Russia
}
\emailAdd{ml@ms2.inr.ac.ru}
\emailAdd{sa.mironov\underline{\ }1@physics.msu.ru}
\emailAdd{rubakov@ms2.inr.ac.ru}
\abstract{
We study spatially flat bouncing cosmologies and models with the
early-time Genesis epoch in a popular class of generalized Galileon
theories. We ask whether there exist solutions of these types which are
free of gradient and ghost instabilities. We find that irrespectively of
the forms of the Lagrangian functions, the bouncing models either are
plagued with these instabilities or have singularities. The same result
holds for the original Genesis model and its variants in which the scale
factor tends to a constant as $t\to -\infty$. The result remains valid in
theories with additional matter that obeys the Null Energy Condition and
interacts with the Galileon only gravitationally. We propose a modified
Genesis model which evades our no-go argument and give an explicit example
of healthy cosmology that connects the modified Genesis epoch with
kination (the epoch still driven by the Galileon field, which is a
conventional massless scalar field at that stage).
}

\begin{document}
\begin{flushright}
INR-TH-2016-014
\end{flushright}
\maketitle
\flushbottom

\section{Introduction and summary}

Bouncing and Genesis cosmologies are interesting scenarios alternative or
complementary to inflation. Both require the violation of the Null Energy
Condition\footnote{An exception is bounce of a closed
Universe~\cite{Starobinsky-bounce}.} (NEC), and hence fairly
unconventional matter. Candidates for the latter are generalized
Galileons~\cite{Horndeski:1974wa,Fairlie:1991qe,Luty:2003vm,Nicolis:2008in,
Deffayet:2010zh,Deffayet:2010qz,Kobayashi:2010cm,Padilla:2012dx}, scalar
fields whose Lagrangians involve second derivatives, and whose field
equations are nevertheless second order (for a review see, e.g.,
Ref.~\cite{Rubakov:2014jja}). Indeed, in the original Genesis
model~\cite{Genesis1} as well as in its
variants~\cite{Genesis2,Hinterbichler:2012fr,Elder:2013gya,Liu:2011ns,
Pirtskhalava:2014esa,Nishi:2015pta}, the initial super-accelerating stage
can occur without ghosts and gradient instabilities (although there is
still an issue of
superluminality~\cite{Easson:2013bda,Kolevatov:2015iqa}). Likewise,
 bouncing Universe models with generalized Galileons can be arranged in
such a way that no ghost or gradient instabilities occur at and near the
bounce~\cite{Qiu:2011cy,Easson:2011zy,Osipov:2013ssa,Qiu:2013eoa}.

The situation is not so bright in more complete cosmological models. Known
models of the bouncing Universe, employing generalized Galileons, are in
fact plagued by the gradient instabilities, provided one follows the
evolution for long enough
time~\cite{Cai:2012va,Koehn:2013upa,Battarra:2014tga,Qiu:2015nha,Koehn:2015vvy}.
Gradient instabilities occur also in the known Genesis models, once one
requires that the early Genesis regime turns into more conventional
expansion (inflationary or not) at later
times~\cite{Pirtskhalava:2014esa,Kobayashi:2015gga,Sosnovikov}. An
intriguing exception is the model~\cite{Pirtskhalava:2014esa} in which
Genesis-like super-accelerated expansion starts from the de~Sitter, rather
than Minkowski, epoch. We comment on this model in
Section~\ref{sec:no-go}.

One way to get around the gradient instability problem is to arrange the
 model in such a way that the quadratic in spatial gradients, wrong sign
term in the action is small, and higher derivative terms restore stability
at sufficiently high spatial
momenta~\cite{Pirtskhalava:2014esa,Creminelli:2006xe}. There is also a
possibility that the strong coupling momentum scale is low
enough~\cite{Koehn:2015vvy}. In both cases  the exponential growth of
trustworthy perturbations does not have catastrophic consequences,
provided that the time interval at which the instability operates is short
enough. Another option is to introduce extra terms in the action which are
not invariant under general coordinate
transformations~\cite{Qiu:2015nha,Kobayashi:2015gga}.

Clearly, it is of interest to understand whether gradient or ghost
instabilities are inherent in all ``complete'' bouncing models and Genesis
models with initial Minkowski space, which are based on classical
generalized Galileons and General Relativity, or these instabilities are
merely drawbacks of concrete models constructed so far. In the latter
case, it is worth designing examples in which the gradient and ghost
instabilities are absent.

It is this set of issues we address in this paper. We consider the
simplest and best studied generalized Galileon theory interacting with
gravity. The Lagrangian is (mostly negative signature;  $\kappa = 8\pi G$)
\be
L =  - \frac{1}{2\kappa}R + F(\pi, X) + K (\pi, X) \Box \pi  \; ,
\label{sep22-15-30}
\ee
where  $\pi$ is the Galileon field, $F$ and $K$ are smooth Lagrangian
functions, and
\[
X = \nabla_\mu \pi \nabla^\mu \pi \; , \;\;\;\; \Box \pi = \nabla_\mu
\nabla^\mu \pi \; .
\]
We also allow for other types of matter, assuming that they interact
with the Galileon only gravitationally and obey the NEC:
\be
\rho_M + p_M \geq 0 \; .
\label{may7-16-20}
\ee
To see that our observations are valid in any dimensions, we study
this theory in $(d+1)$ space-time dimensions with $d \geq 3$; the case of
interest is of course $d=3$. We consider spatially flat FLRW Universe with
the scale factor $a(t)$ where $t$ is the cosmic time, and study spatially
homogeneous backgrounds $\pi(t)$.

Our framework is quite general. In the Genesis case we require that
neither $a(t)$ nor $\pi(t)$ has future singularity (i.e., $a(t)$, $\pi(t)$
 and their derivatives are finite for all $-\infty <t < +\infty$). Our
definition of the bouncing Universe  is that the scale factor $a(t)$
either is constant in the past and  future, $a (t) \to a_{\mp}$ as $t \to
\mp \infty$, or diverges in one or both of the asymptotics (i.e., $a_- =
\infty$ or/and $a_+ = \infty$), and that there is no singularity in
between.

Somewhat surprisingly, our results for the bouncing and Genesis scenarios
are quite different. In the bouncing Universe case, we show that the
gradient (or ghost) instability is inevitable. This result is a
cosmological counterpart of the observation that a static, spherically
symmetric Lorentzian wormhole supported by the generalized Galileon always
has the ghost or gradient instability~\cite{Rubakov:2016zah} (see also
Ref.~\cite{Rubakov:2015gza}); the technicalities involved are also
similar.

Analogous no-go theorem does not hold in the Genesis case. Yet the
requirement of the absence of the gradient and ghost instabilities
strongly constrains the Galileon theories (i.e., Lagrangian functions $F$
and $K$). In particular, the gradient or ghost instability (or future
singularity) does exist, if the initial stage is the original
Genesis~\cite{Genesis1} or its versions in which $a(t) \to \mbox{const}$
as $t \to -\infty$, which is the case, e.g., in the subluminal
Genesis~\cite{Genesis2} as well as in the DBI~\cite{Hinterbichler:2012fr}
and generalized Genesis~\cite{Nishi:2015pta} models in which the
Lagrangians have the general form\footnote{The reservation here has to do
with the fact that we merely leave more complicated models aside. It is
worth seeng whether our result can be generalized to all Horndeski-like
Lagrangians.} \eqref{sep22-15-30} (in the language of
Ref.~\cite{Hinterbichler:2012fr}, the Lagrangians from this subclass do
not contain terms ${\cal L}_4$ and ${\cal L}_5$).

Equipped with better understanding of the instabilities in the Genesis
models with generalized Galileons, we propose a modified Genesis behavior
in which the space-time curvature, energy and pressure vanish as $t \to
-\infty$ and which
is not inconsistent with the absence of the gradient and ghost
instabilities and the absence of future singularity. The pertinent
Galileon Lagrangian is similar to ones considered in
Refs.~\cite{Liu:2011ns,Nishi:2015pta}; in particular, the action is not
scale-invariant. Starting from this Lagrangian, we give an example of a
``complete'' model, with Genesis at the initial stage and kination (the
epoch still driven by the Galileon field which, however, is a conventional
massless scalar field at that stage) at later times. This model is free of
the gradient instabilities, ghosts and superluminal propagation about the
homogeneous solution, while the kination stage may possibly be connected
to the radiation domination epoch via, e.g., gravitational particle
creation, cf. Ref.~\cite{Nishi:2016wty}.

This paper is organized as follows. In Section~\ref{sec:general} we
discuss, in general terms, the conditions for the absence of the gradient
and ghost instabilities in the cosmological setting. We show in
Section~\ref{sec:no-go} that irrespectively of the forms of the Lagrangian
functions $F(\pi, X)$ and $K(\pi, X)$, these conditions cannot be
satisfied in the bouncing Universe scenario as well as in the Genesis
models with time-independent past asymptotics of the scale factor. We
propose a modified Genesis model in Section~\ref{sec:modified}, where we
first study general properties and concrete example of early Genesis-like
epoch which evades the no-go argument of Section~\ref{sec:no-go}, then
give an explicit example of healthy model connecting Genesis and kination
and, finally, briefly discuss a spectator field whose perturbations may
serve as seeds of the adiabatic perturbations. We conclude in
Section~\ref{sec:conclusion}. For completeness, the general expressions
for the Galileon energy-momentum tensor and quadratic Lagrangian of the
Galileon perturbations are given in Appendix.

\section{Generalities}
\label{sec:general}

The general expression for the Galileon energy-momentum tensor is given in
Appendix, eq.~\eqref{may7-16-1}. In the cosmological context
the energy density and pressure are
\begin{subequations}
\label{jan3-16-2}
\begin{align}
\rho &= 2F_X X - F - K_\pi X + 2d H K_X \dot{\pi}^3 \; ,
\label{jan3-16-2a}
\\
p &= F - 2 K_X X \ddot{\pi} - K_\pi X \; ,
\label{jan3-16-2b}
\end{align}
\end{subequations}
where $H$ is the Hubble parameter and
\[
X = \dot{\pi}^2 \; .
\]
Hereafter $F_\pi = \d F/\d \pi$, $F_X = \d F/\d X$, etc.

Our main concern is the Galileon perturbations $\chi$ of high momentum and
frequency. The general expression for the effective quadratic Lagrangian
for perturbations is again given in Appendix, eq.~\eqref{jan3-16-1}. For
homogeneous background we obtain
\be
L^{(2)} = A \dot{\chi}^2 - \frac{1}{a^2} B (\d_i \chi)^2 + \dots
\label{may7-16-30}
\ee
where
\begin{subequations}
\begin{align}
A &=  F_X + 2F_{XX}X - K_\pi - K_{X \pi} X + 2dH \dot{\pi} (K_X + K_{XX}
X) + \frac{2d}{d-1}\kappa K_X^2 X^2 \; , \\
B &= F_X - K_\pi + 2K_X \ddot{\pi} + K_{X\pi}X + 2 K_{XX} X \ddot{\pi}
+ 2(d-1)HK_X \dot{\pi} - \frac{2(d-2)}{d-1} \kappa K_X^2 X^2 \; ,
\label{jan3-16-3}
\end{align}
\end{subequations}
and terms omitted in \eqref{may7-16-30} do not contain second derivatives
of $\chi$. These terms are irrelevant for high momentum modes. The absence
of ghosts and gradient instabilities requires $A>0$, $B \geq 0$. In
particular, if $B<0$, there are ghosts (for $A <0$) or gradient
instability (for $A>0$).

Our focus is on the coefficient $B$. Despite appearance, it can be cast in
a simple form. To this end we make use of the Friedmann and covariant
conservation equations
\begin{subequations}
\label{jan5-16-10}
\begin{align}
H^2 = \frac{2}{d(d-1)} \kappa (\rho + \rho_M) \\
\dot{\rho} = - d\cdot H(\rho+p)
\label{may7-16-40}\\
\dot{\rho}_M = - d\cdot H (\rho_M + p_M)
\label{may7-16-41}
\end{align}
\end{subequations}
and hence
\be
\dot{H} = - \frac{1}{d-1} \kappa [(\rho +p) + (\rho_M + p_M)] \; .
\label{jan3-16-4}
\ee
Here $\rho$ and $p$ are Galileon energy density and pressure, while
$\rho_M$ and $p_M$ are energy and pressure of conventional matter, if any.
The latter obey the NEC, eq.~\eqref{may7-16-20}. We recall that we assume
that conventional matter does not interact with the Galileon directly, so
the covariant conservation equations \eqref{may7-16-40} and
\eqref{may7-16-41} have to be satisfied separately.

Equations \eqref{jan3-16-2}, \eqref{jan3-16-3} and \eqref{jan3-16-4} lead
to a remarkable relation
\[
2BX = \frac{d}{dt} \l 2K_X \dot{\pi}^3 - \frac{d-1}{\kappa} H \r -
\frac{2(d-2)}{d-1} \kappa K_X \dot{\pi}^3 \l 2K_X \dot{\pi}^3 -
\frac{d-1}{\kappa} H \r - (\rho_M+p_M) \; .
\]
It is natural to introduce a combination
\be
Q = 2K_X \dot{\pi}^3 - \frac{d-1}{\kappa} H
\label{may18-16-1}
\ee
and write
\be
2BX = \dot{Q} -\frac{2(d-2)}{d-1} \kappa K_X \dot{\pi}^3 Q
 - (\rho_M+p_M) \; .
\label{jun7-16-1}
\ee
Another representation is in terms  of the function
\[
R = \frac{Q}{a^{d-2}} \; ,
\]
namely
\[
\frac{2BX}{a^{d-2}} =  \dot{R} - \frac{d-2}{d-1} \kappa a^{d-2} R^2 -
\frac{\rho_M + p_M}{a^{d-2}} \; .
\]
%
Since we assume that the conventional matter, if any, obeys the NEC, the
positivity of $B$ requires
\be
\dot{Q} -\frac{2(d-2)}{d-1} \kappa K_X \dot{\pi}^3 Q \geq 0
\label{jan3-16-5}
\ee
and
\be
\dot{R} - \frac{d-2}{d-1} \kappa a^{d-2} R^2 \geq 0 \; .
\label{may8-16-1}
\ee
As we now see, these requirements are prohibitively restrictive in the
bouncing Universe case and place strong constraints on the Genesis models.

\section{Bouncing Universe and original Genesis: no-go}
\label{sec:no-go}

We now show that the inequality \eqref{may8-16-1} cannot be satisfied in
the bouncing Universe scenario. We write it as follows,
\[
\frac{\dot{R}}{R^2} \geq  \frac{d-2}{d-1} \kappa a^{d-2}
\]
and integrate from $t_i$ to $t_f > t_i$:
\be
\frac{1}{R(t_i)} - \frac{1}{R(t_f)} \geq \frac{d-2}{d-1} \kappa
\int_{t_i}^{t_f}~dt~a^{d-2} \; .
\label{may8-16-2}
\ee
Suppose now that $R(t_i) > 0$. Since $\dot{R} >0$ in view of
\eqref{may8-16-1}, $R$ increases in time and remains positive. We have
\be
\frac{1}{R(t_f)} \leq \frac{1}{R(t_i)} - \frac{d-2}{d-1} \kappa
\int_{t_i}^{t_f}~dt~a^{d-2} \; .
\label{may8-16-10}
\ee
Since  $a(t)$ is either a constant or growing function of $t$ at large
$t$, the right hand side of the latter inequality eventually becomes
negative at large $t_f$. Thus $R^{-1} (t_f)$ as function of $t_f$ starts
positive (at $t_f = t_i$) and necessarily crosses zero. At that time
$R^{-1} = 0$, and $R=\infty$, which means a singularity.

A remaining possibility is that $R(t)$ is negative at all times. In
particular, $R(t_f) < 0$. In that case a useful form of the
inequality~\eqref{may8-16-2} is
\be
\frac{1}{R(t_i)} \geq \frac{1}{R(t_f)} +  \frac{d-2}{d-1} \kappa
\int_{t_i}^{t_f}~dt~a^{d-2} \; .
\label{may8-16-11}
\ee
Now,  $a(t)$ is either a constant or tends to infinity as $t \to
-\infty$, so the right hand side is positive at large negative $t_i$.
Hence, there is again a singularity $R =\infty$ at $t_i < t < t_f$. This
completes the argument.

The same argument applies to the original Genesis model~\cite{Genesis1}
and many of its versions, like subluminal Genesis~\cite{Genesis2} and the
DBI Genesis~\cite{Hinterbichler:2012fr}, provided the Lagrangian has the
general form \eqref{sep22-15-30}. In these versions, the scale factor
tends to a constant as $t \to -\infty$ and, assuming that the Universe
ends up in the conventional expansion regime, the scale factor grows at
large times. The integral in eq.~\eqref{may8-16-2} blows up at large $t_f$
or large negative $t_i$, so the inequalities \eqref{may8-16-10},
\eqref{may8-16-11} are impossible to satisfy without hitting the
singularity\footnote{Note that if there is an initial singularity, there
is no argument that would forbid $Q$ to be negative and increasing towards
zero at early times, cross zero at some intermediate time and continue to
increase later on. This is what happens in the setup~\cite{Elder:2013gya}
where the NEC is satisfied at early times and is violated later on in the
Genesis phase. The inequality \eqref{may8-16-10} shows, however, that in
that case there is either gradient (or ghost) instabitity or future
singularity after $Q$ crosses zero.} $R = \infty$. In fact, in the models
of Refs.~~\cite{Genesis1,Genesis2,Hinterbichler:2012fr}, one has $Q>0$,
which is consistent with healthy behavior at early times but implies
either gradient (or ghost) instabitity or singularity in future.

At this point let us make contact with the model of
Ref.~\cite{Pirtskhalava:2014esa} in which the Genesis-like
super-accelerated expansion starts from the de~Sitter rather than
Minkowski epoch, $d=3$, $a \propto \e^{\lambda t}$. In that case the
integral in \eqref{may8-16-11} is convergent as $t_i \to -\infty$. Hence,
our argument does not work: one can have $R<0$ at all times, leaving a
room for the stable evolution. In fact, in the model of
Ref.~\cite{Pirtskhalava:2014esa}, our parameter $Q$ defined in
\eqref{may18-16-1} is constant in time and negative, while $B>0$ in full
accordance with \eqref{jun7-16-1}. We generalize this construction in
Section~\ref{sec:modified}.

\section{Modified Genesis}
\label{sec:modified}

\subsection{Early-time evolution}

In this Section we construct a model which  interpolates between a stage
similar to Genesis (in the sense that space-time curvature, energy and
pressure vanish as $t\to -\infty$) and kination epoch at with the Galileon
behaves as a conventional massless scalar field. The model is purely
classical and does not have gradient or ghost instability at any time. We
begin with the early Genesis-like stage, having in mind the observations
made in Section~\ref{sec:no-go}.

Since we would like the scale factor to increase at late times, and in
view of the inequality \eqref{may8-16-10}, we require that at the
Genesis-like stage $R<0$ and hence
\[
Q < 0 \; .
\]
This means that
\be
H > \frac{2\kappa}{d-1} K_X \dot{\pi}^3 \; .
\label{jan5-16-1}
\ee
Furthermore, the second term in the right hand side of
eq.~\eqref{jan3-16-5} must be larger than $|\dot{Q}|$ at the Genesis-like
epoch: since $H$ increases at that epoch from originally zero value, so
does $|Q|$ (barring cancelations), and we have $\dot{Q} < 0$. Thus,
besides the inequality \eqref{jan5-16-1} we require
\be
\frac{2(d-2)}{d-1}\kappa K_X \dot{\pi}^3 |Q| > |\dot{Q}| \; .
\label{jan5-16-15}
\ee
Assuming power law behavior of $Q$, we see that the simplest option is
that at large negative times
\be
 K_X \dot{\pi}^3 \propto (-t)^{-1} \; , \;\;\;\;\;\; t \to - \infty.
 \label{may9-16-100}
\ee
From eq.~\eqref{jan5-16-1} we deduce that $H$ cannot rapidly tend to
zero as $t \to -\infty$; we can only have
\be
H = - \frac{h}{t} \;  , \;\;\;\;\; a(t) \propto \frac{1}{(-t)^{h}} \;
, \;\;\;\;\;\; h =\mbox{const} \; , \;\;\;\;\;\; t \to - \infty \; ,
\label{may9-16-101}
\ee
in contrast to the conventional Genesis, in which $H \propto
(-t)^{-3}$. Thus {\it both} energy density and pressure should behave like
$t^{-2}$ as $t \to - \infty$ (while in the originl Genesis one has $p
\propto t^{-4}$, $\rho \propto t^{-6}$).

Now, the no-go argument based on \eqref{may8-16-11} is not valid provided
that the integral in the right hand side is convergent at the lower limit
of integration\footnote{For the case of interest $d=3$ (four space-time
dimensions), eq.~\eqref{jun24-16-1} implies that space-time is
past-incomplete in the sense that past-directed null geodesics reach
spatial infinity $a(t) |{\bf x}| = \infty$ at finite value of the affine
parameter (and past-directed time-like geodesics reach spatial infinity in
finite proper time), cf. Ref.~\cite{Borde:2001nh}. We leave this issue
open in this paper.},
\be
\int_{-\infty}^{t}~dt~a^{d-2} < \infty \; .
\label{jun24-16-1}
\ee
Thus, we require that
\be
h > \frac{1}{d-2} \; .
\label{may8-16-20}
\ee
These are the general properties of the Genesis-like stage which is
potentially consistent with the overall healthy dynamics. Note that unlike
in the original Genesis scenario, the scale factor does not tend to a
constant as $t \to -\infty$. Yet the geometry tends to Minkowski at large
negative times in the sense that the space-time curvature tends to zero,
and so do the energy density and pressure.

One way to realize this scenario is to choose the Lagrangian functions in
the following form
\begin{subequations}
\label{jan22-16-4}
\begin{align}
F &= -f^2 (\d \pi)^2 + \alpha_0 \e^{-2\pi} (\d \pi)^4 \\
K &= \beta_0 e^{-2\pi} (\d \pi)^2 \; .
\end{align}
\end{subequations}
The resulting Lagrangian is similar to those introduced in
Ref.~\cite{Nishi:2015pta}, although the particular exponential dependence
that we have in \eqref{jan22-16-4} was not considered there. There is a
solution
\begin{subequations}
\label{jan25-16-10}
\begin{align}
\e^\pi &= - \frac{1}{H_* t} \\
H &= - \frac{h}{t}  \; , \;\;\;\;\;\;\;\;\;\; t<0 \; ,
\end{align}
\end{subequations}
with time-independent $H_*$ and $h$; we relate them to the parameters
$\alpha_0$, $\beta_0$ and $f$ shortly. For this solution
\[
Q = - \frac{1}{t}\l 2 \beta_0 H_*^2 - \frac{d-1}{\kappa} h \r
\]
and
\[
2BX = \frac{1}{ t^2 } \l 2 \beta_0 H_*^2 - \frac{d-1}{\kappa} h \r \l 1
- 2\frac{d-2}{d-1} \kappa \beta_0 H_*^2 \r \; .
\]
Thus, one has $Q<0$ and no gradient instability at early times ($B>0$),
provided that
\be
\frac{d-1}{2\kappa (d-2)} <\beta_0 H_*^2 < \frac{d-1}{2\kappa}h \; .
\label{jan5-16-11}
\ee
Note that with this choice of parameters, the inequality
\eqref{may8-16-20} is satisfied, as it should.

The parameters $H_*$ and $h$ are related to the parameters of the
Lagrangian via the field equations \eqref{jan5-16-10}. Two independent
combinations of these equations are
\begin{subequations}
\label{jan6-16-1}
\begin{align}
2 \alpha_0 H_*^2 &= \frac{d(d-1)}{\kappa} h^2 + \frac{(d-1)}{\kappa} h - 2
\beta_0 H_*^2 - 2d h  \beta_0 H_*^2 \\
f^2 &= \frac{d(d-1)}{\kappa} h^2 + \frac{3(d-1)}{2\kappa} h -  \beta_0
H_*^2 - d h  \beta_0 H_*^2
\end{align}
\end{subequations}
Using these relations and inequalities \eqref{jan5-16-11} one can check
that $f^2 > 0$ and, importantly, the coefficient of the kinetic term in
the Lagrangian for perturbations is positive, $A >0$. The modified Genesis
regime is stable.

\subsection{From Genesis to kination: an example}

A scenario for further evolution is as follows. The function $\pi(t)$ is
monotonous, $\dot{\pi} > 0$. However, the Lagrangian functions $F$ and $K$
depend on $\pi$ and hence on time in a non-trivial way. The variable $Q$
remains negative at all times, but eventually (at $t=t_c$) $\dot{Q}$
changes sign and $Q$ starts to increase towards zero. Choosing $K_X > 0$,
we find that at $t>t_c$ the stability condition $B>0$ is satisfied
trivially,
\[
2BX = \dot{Q} -  \frac{2(d-2)}{d-1} \kappa K_X \dot{\pi}^3 Q > 0  \; ,
\;\;\;\;\; t>t_c
\]
One should make sure, however, that at $t<t_c$ the inequality
\eqref{jan5-16-15} is always satisfied. Another point to check is that
the coefficient of the kinetic term in the Lagrangian for perturbations is
positive, $A >0$, at all times.

To cook up a concrete example of a model in which the Genesis regime is
smoothly connected to kination (the regime at which Galileon is a
conventional massless scalar field dominating the cosmological expansion),
the simplest way is to introduce a field
\[
\phi = \phi (\pi)
\]
in such a way that the solution is
\be
\phi = t \; .
\label{jan6-16-20}
\ee
The general formulas of Sections~\ref{sec:general}, \ref{sec:no-go}
remain valid, with understanding that the Lagrangian functions are now
functions of $\phi$ and $X =(\d \phi)^2$; in particular, the combination
$Q$ is the same as in \eqref{may18-16-1} with $\phi$ substituted for
$\pi$. We choose the Lagrangian functions in the following form:
\begin{align*}
F &= - v(\phi) X + \alpha (\phi) X^2 \; , \\
K &= \beta(\phi) X \; .
\end{align*}
On the solution \eqref{jan6-16-20} one has
\[
F= -v(t) + \alpha (t) \; , \;\;\;\;\; F_X = -v (t) + 2\alpha (t) \; ,
\;\;\;\;\; K=K_X = \beta(t) \; .
\]
One way to proceed is to postulate suitable forms of $Q(t)<0$ and
$\beta(t)$, such that the inequality \eqref{jan5-16-15} is satisfied,
evaluate $H = \frac{\kappa}{d-1} (2\beta -Q)$, reconstruct $v(t)$ and
$\alpha (t)$ from the field equations and then check that $A>0$ at all
times. With the convention \eqref{jan6-16-20}, once $v(t)$, $\alpha (t)$
and $\beta (t)$ are known, the Lagrangian functions are also known,
$v(\phi) = v(t=\phi)$, etc.

As we anticipated in eqs.~\eqref{may9-16-100}, \eqref{may9-16-101}, the
initial behavior is
\begin{align*}
Q &=  - \frac{\hat{q}}{t} \; , \;\;\;\;\;\; \hat{q} < 0 \\
\beta &= - \frac{\hat{\beta}}{t} \; , \;\;\;\;\;\;\;\; t \to -\infty
\end{align*}
with time-independent $\hat{q}$ and $\hat{\beta}$. To satisfy the
inequality \eqref{jan3-16-5}, we impose the condition
\[
\hat{\beta}> \frac{d-1}{2\kappa (d-2)} \; .
\]
We would like the Galileon to become a conventional massless scalar
field at large positive $t$, whose equation of state is $p=\rho$, and
require that at large $t$ the function $\beta(t)$ rapidly vanishes, while
$H=(d\cdot t)^{-1}$, and hence
\[
Q = - \frac{d-1}{d\kappa t} \; , \;\;\;\;\;\;\;\; t \to +\infty \; .
\]

It is convenient to introduce rescaled variables
\begin{align*}
Q &=  - \frac{{d-1}}{d\kappa} P \; , \\
 \beta &= \frac{{d-1}}{d\kappa} b \; ,
 \end{align*}
where $P$ is positive. In terms of these variables the Hubble parameter is
\[
H = \frac{\kappa}{d-1} (2 K_X \dot{\phi}^3 -Q)= \frac{1}{d}(2b+P) \; .
\]
The combinations of the field equations, analogous to
eqs.~\eqref{jan6-16-1}, give
\begin{subequations}
\label{jan2-16-3}
\begin{align}
\alpha &=  - dH\beta + \frac{d(d-1)}{2\kappa} H^2 + \frac{d-1}{2\kappa}
\dot{H} \nonumber \\
&= \frac{d-1}{d\kappa} \l bP + \frac{1}{2} P^2 + \dot{b} +
\frac{1}{2}\dot{P}\r \; ,\\
v &= -\dot{\beta} - dH\beta + \frac{d(d-1)}{\kappa} H^2 +
\frac{3(d-1)}{2\kappa} \dot{H} \nonumber \\
 &= \frac{{d-1}}{d\kappa} \l 2b^2 + 3 bP +P^2 + 2\dot{b} + \frac{3}{2}
\dot{P} \r \; .
 \end{align}
\end{subequations}
Finally, the coefficients in the Lagrangian for perturbations are
\begin{align*}
A &= \frac{{d-1}}{d\kappa} \l 4b^2 + 5bP + 2P^2 + 2\dot{b} + \frac{3}{2}
\dot{P} \r \; , \\
B &= \frac{{d-1}}{d\kappa} \l \frac{d-2}{d} bP - \frac{1}{2} \dot{P} \r
\; .
\end{align*}
The asymptotics of the solution should be
\begin{subequations}
\label{jan6-16-10all}
\begin{align}
t \to -\infty\;: \;\;\;\;\; P &=    - \frac{p_0}{t} \; , \;\;\;\; p_0 >0
\; , \nonumber \\
b &=  - \frac{b_0}{t} \; , \;\;\;\; b_0 > \frac{d}{2(d-2)} \; ,
\label{jan22-16-2}\\
t \to +\infty\;: \;\;\;\;\; P &= \frac{1}{t} \; , \nonumber \\
b &= 0 \; .
\label{jan6-16-10}
\end{align}
\end{subequations}
As a cross check, the late-time asymptotics \eqref{jan6-16-10} imply
$\alpha = 0$ and
\[
v = - \frac{d-1}{2d\kappa} \cdot \frac{1}{ t^2} < 0 \; ,
\]
which corresponds to the Lagrangian of free massless scalar field,
albeit written in somewhat unconventional form,
\be
L_\phi^{t \to +\infty} = \frac{d-1}{2d\kappa} \frac{(\d
\phi)^2}{\phi^2} \; .
\label{aug8-16-1}
\ee
On the other hand, the early-time asymptotics \eqref{jan22-16-2},
according to eq.~\eqref{jan2-16-3}, give
\begin{align*}
\alpha = \frac{c_1}{t^2} \; , \;\;\;\;\;\; c_1 > 0 \; , \\
v = \frac{c_2}{t^2} \; , \;\;\;\;\;\; c_2 > 0 \; ,
\end{align*}
so that the Lagrangian at early times ($\phi \to -\infty$) reads
\[
L_\phi^{t \to -\infty} = - \frac{c_1}{\phi^2} (\d \phi)^2 +
\frac{c_2}{\phi^2} (\d \phi)^4 - \frac{\hat{\beta}}{\phi} (\d \phi)^2 \Box
\phi \; .
\]
Upon introducing new field at early times
\[
\pi = - \ln(- \phi) \; ,
\]
one writes the Lagrangian in the  following form
\[
L_\pi^{t \to -\infty} = - c_1 (\d \pi)^2 + (c_2 - \hat{\beta})
\e^{-2\pi} (\d \pi)^4 + \hat{\beta} \e^{-2\pi} (\d \pi)^2 \Box \pi \; .
\]
This is precisely the form \eqref{jan22-16-4}.

One more point to note is that if the parametric form of $b (t)$ and
$P(t)$ is
\[
b = \tau^{-1} \tilde{b}(t/\tau)\; , \;\;\;\;\;\; P =\tau^{-1}
\tilde{P}(t/\tau) \; ,
\]
with $\tilde{b}$ and $\tilde{P}$ of order 1 (which is consistent with
the asymptotics \eqref{jan6-16-10all}), then for large $\tau$ the dynamics
is sub-Planckian during entire evolution: in that case one has
sub-Planckian $H \sim \tau^{-1}$,~ $\alpha  \sim \kappa^{-1} \tau^{-2}$,
etc.

\begin{figure}[!tb]
\centerline{\includegraphics[width=0.35\textwidth, angle=-90]{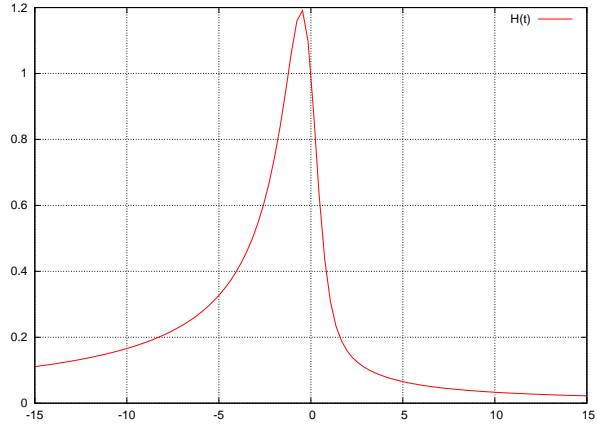}
}
\caption{Hubble parameter in units of $\tau^{-1}$ as function of $t/\tau$
for the model of eqs.~\eqref{jan6-16-30}, \eqref{jan6-16-31}.
\label{Hubble}
}
\end{figure}

\begin{figure}[!tb]
\centerline{\includegraphics[width=0.3\textwidth, angle=-90]{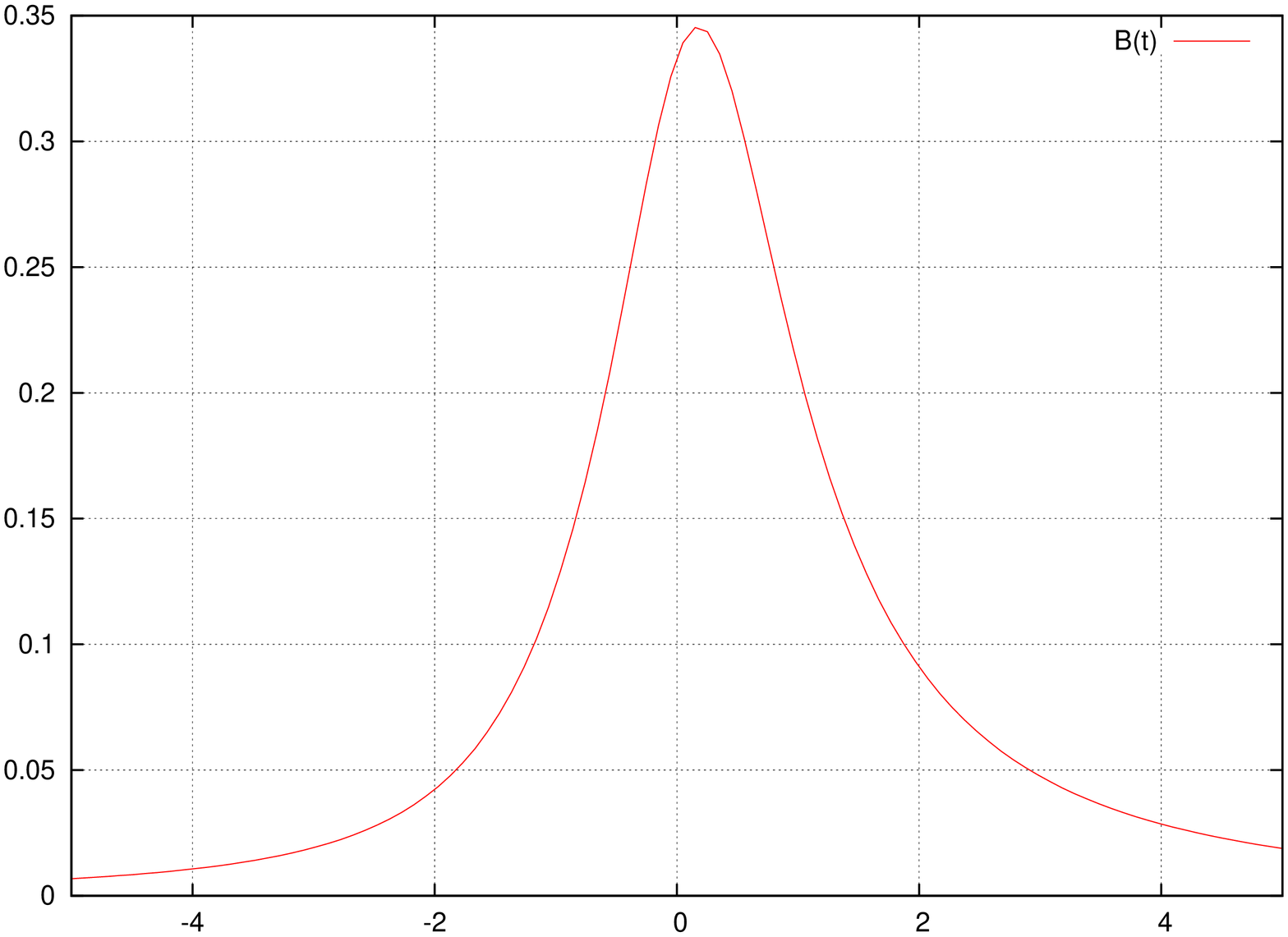}
\includegraphics[width=0.3\textwidth, angle=-90]{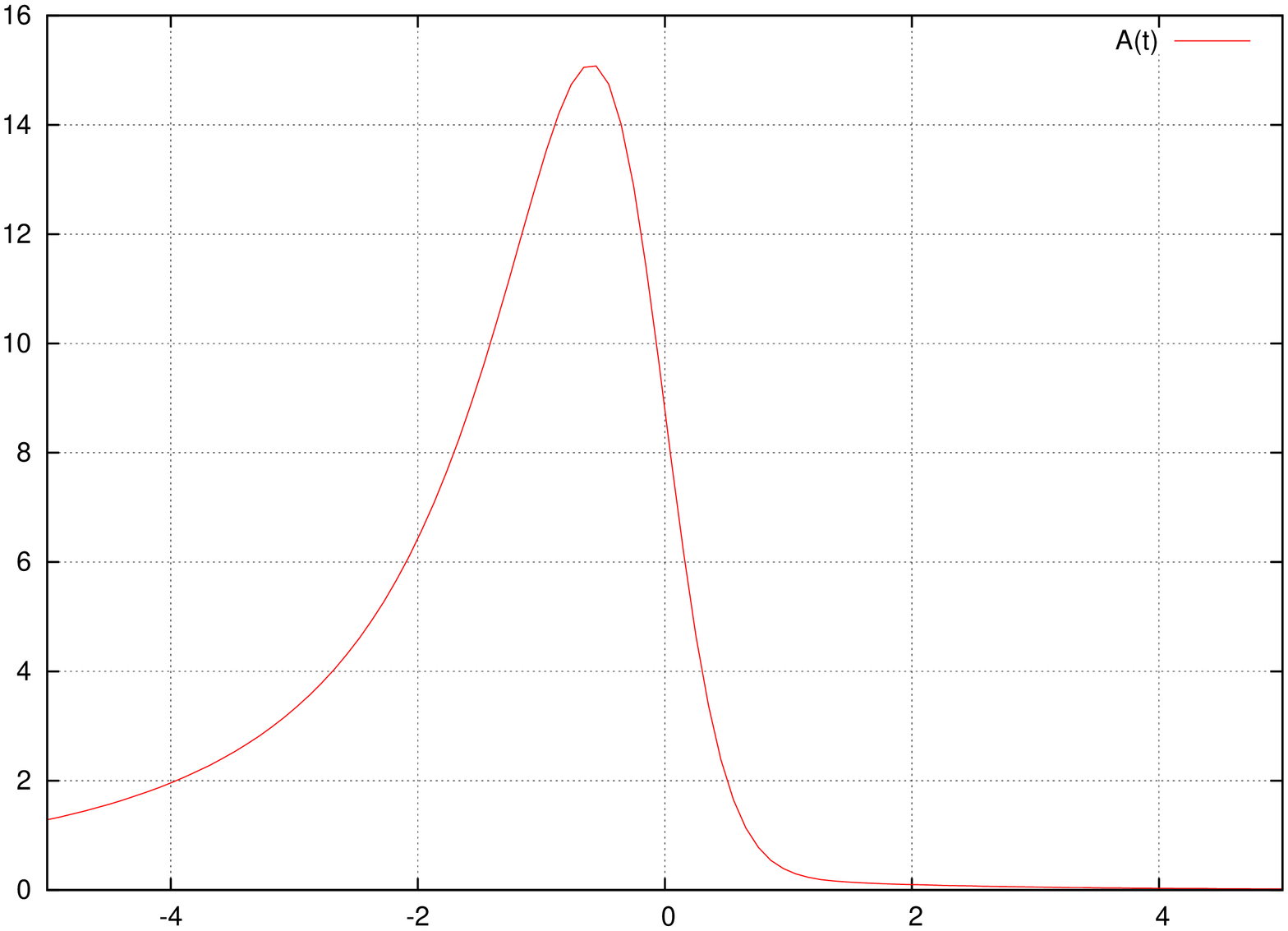}}
\caption{Left: the function $B$ in units $\frac{2}{3\kappa}\tau^{-2}$ as
function of $t/\tau$ for the model of eqs.~\eqref{jan6-16-30},
\eqref{jan6-16-31}. Right: same for the function $A$. Note the larger
scale as compared to the left panel, which implies small sound speed at
early times.
\label{B}
}
\end{figure}

\begin{figure}[!tb]
\centerline{\includegraphics[width=0.35\textwidth, angle=-90]{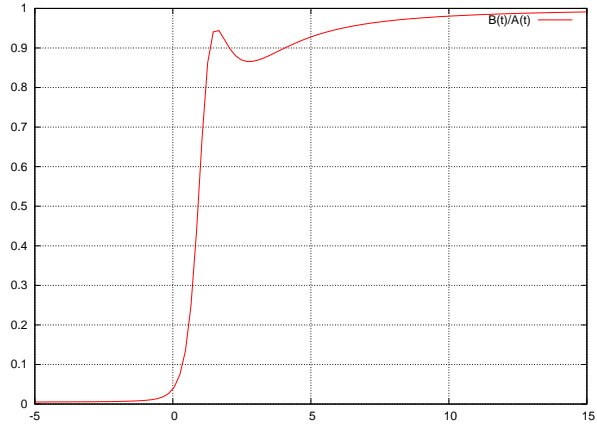}
}
\caption{Sound speed squared, $B/A$, as function of $t/\tau$ for the model
of eqs.~\eqref{jan6-16-30}, \eqref{jan6-16-31}.
\label{sound}
}
\end{figure}

\begin{figure}[!tb]
\centerline{\includegraphics[width=0.30\textwidth, angle=-90]{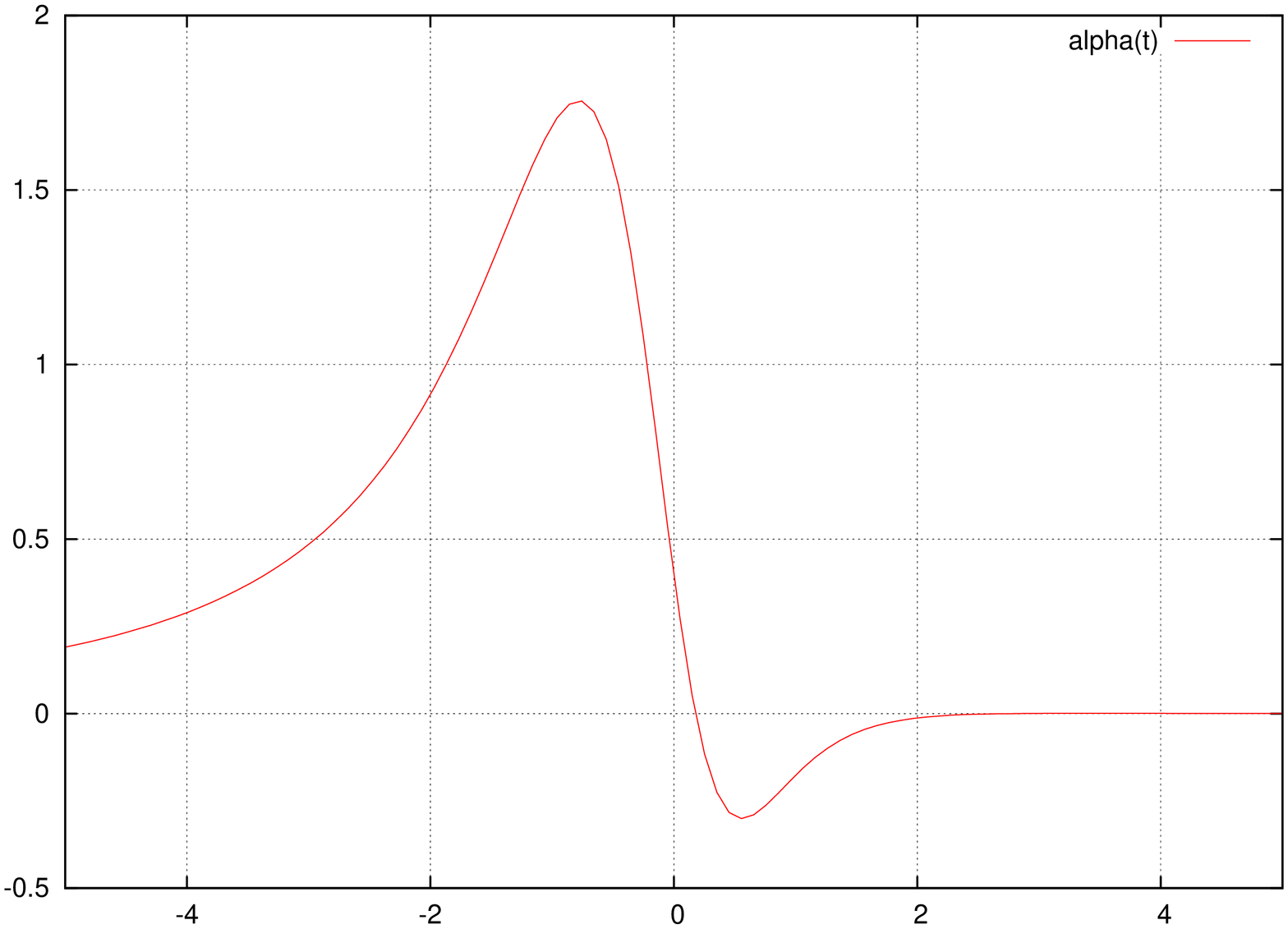}
\includegraphics[width=0.30\textwidth, angle=-90]{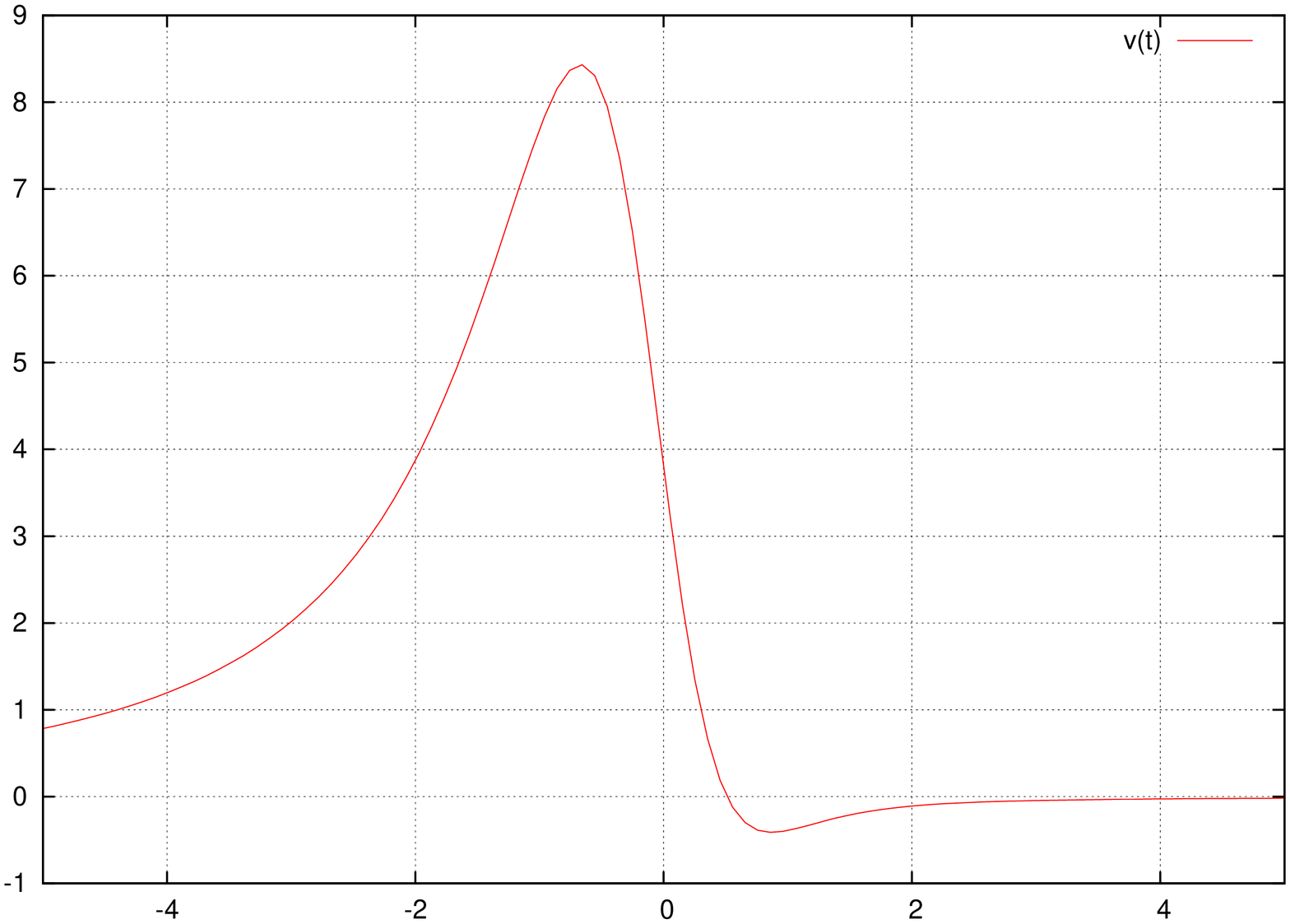}}
\caption{Left: the function $\alpha$ in units $\frac{2}{3\kappa}\tau^{-2}$
as function of $t/\tau$ for the model of eqs.~\eqref{jan6-16-30},
\eqref{jan6-16-31}. Note that $\alpha$ rapidly vanishes at late times.
Right: the same for the function $v$. Negative sign of $v$ at late times
corresponds to conventional sign of the scalar kinetic term.
\label{v}
}
\end{figure}

A random example is
\begin{subequations}
\label{jan6-16-30}
\begin{align}
P &= \frac{1}{\sqrt{t^2 + \tau^2}} \; , \\
b &= \frac{b_0}{\sqrt{t^2 + \tau^2}} \cdot \frac{1}{2}\left[ 1 -
\tanh\l \mu \frac{t}{\tau} \r \right] \; .
\end{align}
\end{subequations}
With
\be
d=3 \; , \;\;\;\; b_0=2 \; , \;\;\;\; \mu=1.1
\label{jan6-16-31}
\ee
($(3+1)$-dimensional space-time) the system smoothly evolves  from the
modified Genesis regime to kination (free massless scalar field) regime
with $A(t) > 0$ and $B(t) > 0$ for all $t$, and also with subluminal
propagation of perturbations about this background (the latter property
 explains the choice $\mu=1.1$: for $\mu=1$, say, there is a brief time
interval in which the perturbations propagate superluminally). The
properties of this model are illustrated in Figs.~\ref{Hubble} -- \ref{v}.
It is worth noting that although $A(t)$ and $B(t)$ at late times
appear different
in Fig.~\ref{B}, they are actually the same,
\[
A(t)= B(t) = \frac{1}{3\kappa t^2} \; , \;\;\;\;\; t \to +\infty \; .
\]
The fact that these functions tend to zero as $t \to +\infty$ does not
indicate the onset of strong coupling; this behavior rather has to do
with the field redefinition from the canonical massless scalar field to
the field $\phi$ described by the Lagrangian \eqref{aug8-16-1}. The 
late-time theory is the theory of free massless scalar field, with no strong
coupling or instabilities.



\subsection{Curvaton}

It is unlikely that the Galileon perturbations would produce adiabatic
 perturbations with nearly flat power spectrum. Like in
Ref.~\cite{Genesis1}, the adiabatic perturbations may originate from
perturbations of an additional scalar field, ``curvaton''. Let us consider
this point, specifying to 4-dimensional space-time, $d=3$.

We are interested in the early modified Genesis stage when the Galileon
Lagrangian functions have the form \eqref{jan22-16-4}. The modified
Galileon action is not invariant under the scale transformations
\[
\pi (x) \to \hat{\pi} (x) = \pi (\lambda x) + \ln \lambda \;
,\;\;\;\;\; g_{\mu \nu} (x) \to \hat{g}_{\mu \nu} = g_{\mu \nu} (\lambda
x) \;.
\]
One has instead $S(\hat{\pi} , \hat{g}_{\mu \nu}) = \lambda^{-2} S(\pi,
g_{\mu \nu})$, just like for the Einstein--Hilbert action. Let us
introduce a spectator curvaton field $\theta$ which is invariant under the
scale transformations, $\theta (x) \to \hat{\theta} (x) = \theta (\lambda
x)$, and require that its action be scale-invariant. This requirement
gives
\[
S_\theta = \int~d^4x~\sqrt{-g} \e^{2\pi} (\d \theta)^2 \; .
\]
In the backround \eqref{jan25-16-10} this action reads
\[
S_\theta = \int~dt d^3x~a^3 \l \frac{1}{(H_* t)^2} \l \frac{\d
\theta}{\d t} \r^2 - \frac{1}{(H_* t)^2} \frac{1}{a^2} \l \frac{\d
\theta}{\d x_i} \r^2 \r \; ,
\]
where the scale factor is given by \eqref{may9-16-101}. Let us
introduce conformal time
\[
\eta = \int~\frac{dt}{a(t)} = - \frac{1}{a_0 (h+1)} (-t)^{h+1} \; .
\]
Then the action for $\theta$ has the form
\[
S_{\theta} = \int~d\eta d^3 x ~a_{eff}^2 (\eta) \left[ \l \frac{\d
\theta}{\d \eta} \r^2 - \l \frac{\d \theta}{\d x_i} \r^2 \right] \; ,
\]
where
\[
a_{eff}(\eta) = - \frac{1}{H_* (h+1) \eta} \; ,
\]
which is precisely the action of the massless scalar field in de~Sitter
space-time. We immediately deduce that the power spectrum of perturbations
$\delta \theta$ generated at the modfied Genesis epoch is flat. This is a
pre-requisite for the nearly flat power spectrum of adiabatic
perturbations, which may be generated from the curvaton perturbations at
later stage.

\section{Conclusion}
\label{sec:conclusion}

We have seen in this paper that with generalized Galileons, it is possible
to construct healthy Genesis-like cosmologies, albeit with somewhat
different properties as compared to the original Genesis scenario. On the
other hand, bouncing cosmologies with generalized Galileons are plagued by
the gradient (or ghost) instability, at least at the level of second
derivative Lagrangians of the form \eqref{sep22-15-30}. This is not
particularly surprising. The theory appears to protect
itself~\cite{Rubakov:2016zah} from having stable wormhole solutions which
can be converted into time machines~\cite{Morris:1988tu}. Technically the
same protection mechanism, with radial coordinate and time interchanged,
forbids the existence of spatially flat bouncing cosmologies. It would be
interesting to understand how general are these features.

\acknowledgments
The authors are indebted to R.~Kolevatov, A.~Sosnovikov, A.~Starobinsky
and A.~Vikman for useful discussions and J.-L.~Lehners and D.~Pirtskhalava
for helpful correspondence. This work has been supported by Russian
Science Foundation grant 14-22-00161.

\appendix
\section{Galileon and its perturbations}

The Galileon energy-momentum tensor in the theory with the
Lagrangian~\eqref{sep22-15-30} reads
\be
T_{\mu \nu} = 2 F_X \d_\mu \pi \d_\nu \pi +  2 K_X \Box \pi \cdot
\d_\mu \pi \d_\nu \pi - \d_\mu K \d_\nu \pi - \d_\nu K \d_\mu \pi - g_{\mu
\nu} F + g_{\mu \nu} g^{\lambda \rho} \d_\lambda K \d_\rho \pi \; ,
\label{may7-16-1}
\ee

We now consider Galileon perturbations, write $\pi = \pi_c  + \chi$, and
 omit subscript $c$ in what follows. We are interested in high momentum
and frequency modes, so we concentrate on terms involving $\nabla_\mu \chi
\nabla_\nu \chi$ in the quadratic effective Largangian or, equivalently,
second order terms in the linearized field equation. A subtlety here is
that the Galileon field equation involves the second derivatives of
metric, and the Einstein equations involve the second derivatives of the
Galileon~\cite{Deffayet:2010qz}, and so do the linearized equations for
perturbations. The trick is to integrate the metric perturbations out of
the Galileon field equation by making use of the Einstein
equations~\cite{Deffayet:2010qz}. In the cosmological setting this is
equivalent to the approach adopted in Ref.~\cite{Kobayashi:2010cm}.

To derive the quadratic Lagrangian for the Galileon perturbations, one
writes the full Galileon field equation
\begin{align*}
 \left( -2F_X + 2 K_\pi - 2K_{X \pi} \nabla_\mu \pi \nabla^\mu \pi  - 2
 K_X \Box \pi \right) \Box \pi   + \left(-4 F_{XX} + 4 K_{X \pi} \right)
\nabla^\mu \pi \nabla^\nu \pi \nabla_\mu \nabla_\nu \pi & \nonumber
\\
 - 4 K_{XX} \nabla^\mu \pi \nabla^\nu \pi \nabla_\mu \nabla_\nu \pi
\Box \pi  + 4 K_{XX} \nabla^\nu \pi \nabla^\lambda \pi \nabla_\mu
 \nabla_\nu \pi \nabla^\mu \nabla_\lambda \pi + 2 K_X \nabla^\mu
\nabla^\nu \pi \nabla_\mu \nabla_\nu \pi & \nonumber
\\
+ 2 K_X R_{\mu \nu} \nabla^\mu \pi \nabla^\nu \pi + \ldots = 0 \; ; &
\end{align*}
hereafter dots denote terms without second derivatives. The subtle term is
the last one here.

The linearized equation can be written in the following form
\begin{align}
 -2 [F_X  + K_X \Box \pi - K_\pi +  \nabla_\nu (K_X \nabla^\nu \pi)]
 \nabla_\mu \nabla^\mu \chi  & \nonumber \\
-2  [2 (F_{XX} + K_{XX} \Box \pi) \nabla^\mu \pi \nabla^\nu \pi -
 2(\nabla^\mu K_X) \nabla^\nu \pi - 2K_X \nabla^\mu \nabla^\nu \pi]
\nabla_\mu \nabla_\nu \chi & \nonumber\\
+ 2 K_X R_{\mu \nu}^{(1)} \nabla^\mu \pi \nabla^\nu \pi + \ldots &= 0
\; ,
\label{oct23-15-1}
 \end{align}
 where the terms without the second derivatives of $\chi$ are omitted, and
 $R_{\mu \nu}^{(1)}$ is linear in metric perturbations. We now make use of
the Einstein equations $R_{\mu \nu} - \frac{1}{2} g_{\mu \nu}R = \kappa
T_{\mu \nu}$, or
\[
R_{\mu \nu} = \kappa \l T_{\mu \nu} - \frac{1}{d-1} g_{\mu \nu}
T^\lambda_\lambda \r \; ,
\]
linearize the energy-momentum tensor and obtain for the last term in
eq.~\eqref{oct23-15-1}
\be
2 K_X R_{\mu \nu}^{(1)} \nabla^\mu \pi \nabla^\nu \pi = - 2 \kappa
K_X^2 \left[-\frac{2(d-2)}{d-1} X^2 \Box \chi + 4X \nabla^\mu \pi
\nabla^\nu \pi \nabla_\mu \nabla_\nu \chi \right] + \ldots \; .
\nonumber
\ee
The resulting linearized Galileon field equation is obtained from the
following quadratic Lagrangian:
\begin{align}
L^{(2)} &= [F_X  + K_X \Box \pi - K_\pi + \nabla_\nu (K_X \nabla^\nu \pi)]
\nabla_\mu \chi \nabla^\mu \chi \nonumber \\
 &+ [2 (F_{XX} + K_{XX} \Box \pi) \nabla^\mu \pi \nabla^\nu \pi -
2(\nabla^\mu K_X) \nabla^\nu \pi - 2K_X \nabla^\mu \nabla^\nu \pi]
\nabla_\mu \chi \nabla_\nu \chi
 \nonumber\\
& -\frac{2(d-2)}{d-1} \kappa K_X^2 X^2 \nabla_\mu \chi \nabla^\mu \chi
+ 4\kappa K_X^2 X \nabla^\mu \pi \nabla^\nu \pi \nabla_\mu \chi \nabla_\nu
\chi \; .
\label{jan3-16-1}
\end{align}
We specify to spatially homogeneous Galileon background in
Section~\ref{sec:general}.

\end{document}